\documentclass[a4paper,oneside]{article}
\usepackage{microtype}
\usepackage[sc]{mathpazo}
\usepackage[T1]{fontenc}
\usepackage[DIV=10,BCOR=0cm,headinclude=true]{typearea}

\usepackage{amsfonts,amsmath,amssymb,amsthm}

\usepackage{graphicx}
\usepackage{float}

\usepackage[british]{babel}
\usepackage{bm}
\usepackage{color}
\usepackage[dvipsnames]{xcolor}
\usepackage[utf8]{inputenc}
\usepackage{bm}
\usepackage{pdflscape}
\usepackage[toc,page]{appendix}
\usepackage[makeroom]{cancel}
\usepackage{caption,subcaption}
\usepackage{multirow}
\usepackage{setspace}
\usepackage{hyperref}
\usepackage{soul}
\usepackage[font=small,labelfont=bf]{caption}
\usepackage{MnSymbol}
\usepackage{relsize}
\usepackage{bigints}
\usepackage{dutchcal}
\usepackage{bbm}

\hypersetup{colorlinks,citecolor=black,filecolor=black,linkcolor=black,urlcolor=black}

\definecolor{refkey}{rgb}{1,0.5,0.5}
\definecolor{labelkey}{rgb}{0.5,1,0.5}

\newcommand{\nc}{\newcommand}
\nc{\nn}{\nonumber}

\usepackage{authblk}

\nc{\D}{\mathrm{d}}
\nc{\Do}{\partial}
\nc{\e}{\mathcal{E}}
\nc{\half}{\frac{1}{2}}
\nc{\vecr}{\mathbf{r}}
\nc{\vecpot}{\mathbf{A}}
\nc{\veloc}{\mathbf{v}}
\nc{\elecf}{\mathbf{E}}
\nc{\kvec}{\mathbf{k}}
\nc{\rhovec}{\bm{\rho}}
\nc{\cF}{{\cal F}}

\def\be{\begin{equation}}
\def\ee{\end{equation}}

\usepackage[normalem]{ulem}
\newcommand{\E}{{\cal E}}

\definecolor{mymaroon}{RGB}{148,0,0}
\definecolor{myred}{RGB}{255,126,121}

\title{Intraband vs. interband optical nonlinearities of transparent conducting oxides}
\author{Yonatan Sivan$^{1\ast}$, Nathaniel Kinsey$^2$\\
\normalsize{$^{1}$School of Electrical and Computer engineering, Ben-Gurion University of the Negev, P.O. Box 653, Israel, 8410501}\\
\normalsize{$^{2}$}Department of Electrical and Computer Engineering, Saint Louis University, Saint Louis, Missouri, 63103, USA}

\date{\today}

\begin{document}

\maketitle

\begin{abstract}
We show that the intraband thermo-optical response to ultrashort intense infra-red illumination of transparent conducting oxides is much stronger than the nonlinear response associated with resonant and non-resonant interband transitions in typical experimental setups. This finding rules out several speculative explanations of recent measurements.  
\end{abstract}

\section{Introduction}
One of the ways to classify nonlinear optical effects is via their speed and strength~\cite{Khurgin_Kinsey_LPR,nlty_review,Khurgin-review-2023,NK-review-2024}. In particular, the abundant Kerr effect~\cite{Boyd-book,Fibich-book} is known to be (nearly) instantaneous yet weak; this stems from its non-resonant nature. Indeed, it involves virtual transitions of valence electrons to the conduction band in solids. Wave mixing interactions have similar characteristics. On the other hand, resonant (i.e., real) transitions gives rise to much stronger nonlinear effects, but while they modify the permittivity instantaneously, their decay time is limited by the lifetime of the excited electron states. The latter class of nonlinear effects include free carrier generation due to valence to conduction band transitions, involving resonant single/multi photon absorption, thermal effects, stress, strain etc.. Thus, the scientific community is in an ongoing quest to find the nonlinear optical material that allows enjoying the best of the two worlds - a nonlinear optical effect which is both fast and strong. % we may not always want instantaneous effects. Perhaps simply saying 'fast and strong' is sufficient. The user defines how fast. 

On this background, there was a great deal of excitement when the first studies of transparent conducting oxides (TCOs) showed a unusually strong and fast nonlinear optical response~\cite{Kinsey-Optica-2015, Boyd_NLO_ENZ_ITO,Shalaev_Faccio_NLO_ENZ}. Those materials are highly doped semiconductors; they are also occasionally referred to as low electron density metals~\cite{Un-Sarkar-Sivan-LEDD-I} or $p$-doped metals~\cite{Zayats_CuS}. Originally, the strong nonlinear optical response of TCOs was associated with the proximity to the frequency at which the permittivity of the TCO changes sign from positive to negative; this is the well-known epsilon-near-zero (ENZ) point. In this spectral regime, even though the {\em absolute} change of the permittivity may not be much different from other spectral regimes, the {\em relative} permittivity change can, ideally, diverge. Indeed, it is the relative change which is responsible to the strength of the nonlinear response~\cite{Khurgin-diffusive-switching, Kinsey-Optica-2015,Gurwich-Sivan-CW-nlty-metal_NP}. %{\bf NK, I know of two specific papers that show this , but surely there are more suitable, earlier papers that show this. any come to mind? \color{red} NK says - the argument is poorly formed and not the main point, but it was made in Kinsey-Optica-2015. This is the first i know of it said in this way. However, i would argue that the statement depends on what you want to achieve. If its R,T control, then its the variation in the phase accumulation which is more index related. Perhaps it can be rephrased to say 'relative change in constituent parameters' just to allow a focus on permittivity or index depending upon the focus?}
This observation motivated a wide range of nonlinear optical effects in TCOs, all showing indeed unusually strong effects~\cite{Capretti-Engheta-2015, Clerici_Nat_Comm_2017,Yang-HHG-CdO-natphys-2019,Sapienza_Shalaev_2020, Exeter_Nat_Comm_2021, Ellenbogen-Minerbi-ITO,Sapienza_2022,ENZ_FROG_2022}, see the following reviews for more information~\cite{Boyd_NLO_ENZ_analysis, Kinsey-Nat-Rev-Mater-2019,Capretti-Engheta-2015}.

However, it was soon realized that the relative permittivity change can never diverge because of the small, yet non-zero imaginary part of the permittivity. Thus, the response has ``only'' a resonance at the ENZ point~\cite{Boyd_slow_light_review,Khurgin_slow_light_review}. % left out:Harris: (https://doi.org/10.1103/PhysRevLett.101.103601)}. 
Instead, the strong nonlinear effects seen in measurements were shown to originate from the non-parabolicity of the conduction band~\cite{Kane-quasilinear}, and associated dependence of the plasma frequency~\cite{Boyd_Nat_Phot_2018} on the electron temperature\footnote{A temperature dependence was assumed already in~\cite{Boyd_NLO_ENZ_ITO} in the context of the chemical potential. Later work revealed that this dependence is significant, but not as much as the temperature dependence of the plasma frequency. }. Later work re-framed this dependence in the (more transparent) context of the momentum/energy dependence of the effective mass of the band structure of semiconductors~\cite{Kinsey_ENZ_OptMatExp}. The analysis in~\cite{Kinsey_ENZ_OptMatExp} pointed to the total absorbed energy as the key parameter affecting the temperature rise, hence, the permittivity change; hence, it also indicates its instantaneous absorptive nature. 

% Further studies %{\bf NK - who was the first to do that? Boyd 2018? or later?} {\color{red} NK - well in the ENZ regime it would be Boyd's science paper in 2016 so far as i know. then our work in PRL in 2016 and Nat Comm. in 2017. But this effect well known from metals as you note below, often termed Fermi Smearing nonlinearity. So we can say it was applied to TCOs in 2016, but it was by no means new. --- not accurate; what people called Fermi smearing had to do with the effect of heating on interband transitions. } % associated the permittivity change with the rise of the electron temperature. 

This insight raised the need to determine the temperature. This originally was done using the Two Temperature Model (TTM)~\cite{Two_temp_model,non_eq_model_Ippen,delFatti_nonequilib_2000,Stoll_review}, as done earlier in noble metals. However, the formulae used for the thermo-physical parameters in the TTM did not account for the unique features of the TCOs, namely, the low conduction electron density and the non-parabolicity. As a consequence, the TTM-based theory provided a qualitative match to measurements, but not a quantitative one~\cite{Clerici_Nat_Comm_2017,de_Leon_Mexicans_2022}. % add the Boyd 2018 Nat Phot.? i'd leave out as it is metastructure so it there are other effects at play.

A remedy to this limitation was provided by Sarkar, Un {\em et al.} in~\cite{Un-Sarkar-Sivan-LEDD-I}, who derived the thermophysical parameters of TCOs rigorously from a Boltzmann model for the electron dynamics. This work showed that the electron heat capacity of TCOs is much smaller than for noble metals% {\bf NK - anyone noted this before us? NK says - not that i'm aware of. It was suspected but not shown.}
, and should include high-order corrections due to the elevated temperature, as in textbook formulae~\cite{Ashcroft-Mermin}. More importantly, it showed that a correct account for the $e-ph$ interactions requires accounting also for momentum conservation (this was not necessary for noble metals). With these considerations, and with a Lindhard-based model for the permittivity~\cite{Un-Sarkar-Sivan-LEDD-II}, a quantitative match to experiments was obtained~\cite{de_Leon_Mexicans_2022}. Remarkably, the latter work also showed that a thermal model (i.e., based on the TTM, which naturally emerges from the microscopic Boltzmann model for the electron distribution function~\cite{delFatti_nonequilib_2000,Un-Sarkar-Sivan-LEDD-I} by multiplying by the energy and summing over all energy states) yields almost the same predictions as the non-thermal model. It showed also that there is no need to include the non-thermal part appearing in the so-called extended two temperature model~\cite{non_eq_model_Carpene,Sivan_Spector_metal_diffusion} even for pulses only a few femtosecond long. All this happens because the time scale for $e-e$ induced thermalization is faster than even for noble metals~\cite{Khurgin_Kinsey_LPR,Un-Sarkar-Sivan-LEDD-I}. This model thus reveals a simple physical picture - the permittivity changes essentially instantaneously by absorption of light - it is fairly insensitive to whether the electron distribution is thermal or non-thermal; it then relaxes on the cooling rate, determined by the strength of $e-ph$ interactions. This shows that the nonlinearity associated with conduction band electrons is essentially {\em thermo-optic}. Further studies based on the more advanced, fully-quantum coherent density matrix formulation~\cite{single_cycle_nlty_Article}, demonstrated the small role of the deviation from equilibrium also for pulses no more than a few femtosecond long.

At about the same time, a new gap opened between the TCO theory and measurements. In particular, in the quest for (the already demonstrated) time reversal of optical pulses~\cite{Marom_Fainman,Kuzucu_Gaeta} and temporal photonic crystals at optical frequencies, Lustig {\em et al.} compressed their pump pulse to a single cycle and TW/cm$^2$ intensities, and measured the resulting changes to probe transmissivity and reflectivity from a thin ITO layer~\cite{Shalaev_Segev_nanophotonics_2023}. Surprisingly, they observed a probe limited permittivity rise and decrease in transmissivity, which nevertheless was replaced midway by a very slow decay; the reflectivity showed the usual fast rise-slow decay dynamics. Peculiarly, this feature was  observed in a more recent experiment only with a much longer and much more intense pump pulse. Complex dynamics was observed also in~\cite{Purdue-oscilatory_nlty_ITO-2026}, measured with an even longer probe pulse. None of these complex dynamical features was reproduced in the density-matrix based model of~\cite{single_cycle_nlty_Article,single_cycle_nlty_Letter}, which included only the conduction electrons. Several recent works proposed that nonlinearities associated with interband transitions may be able to explain the complex dynamics observed in these experiments~\cite{Narimanov_2025,Scalora-ITO-2025,Purdue-oscilatory_nlty_ITO-2026,Ferrera-Scalora-ITO-2026,Narimanov_Shapiro_2026}. 

% {\bf Due to the extremely fast $e-e$ interactions in TCOs, the TTM yields essentially the same permittivity dynamics as obtained from the BE model~\cite{Un-Sarkar-Sivan-LEDD-II}, as well as from an even more accurate density matrix model~\cite{single_cycle_nlty_Article,single_cycle_nlty_Letter}.}

\begin{figure}[t]
\centering
\includegraphics[width=0.4\columnwidth]{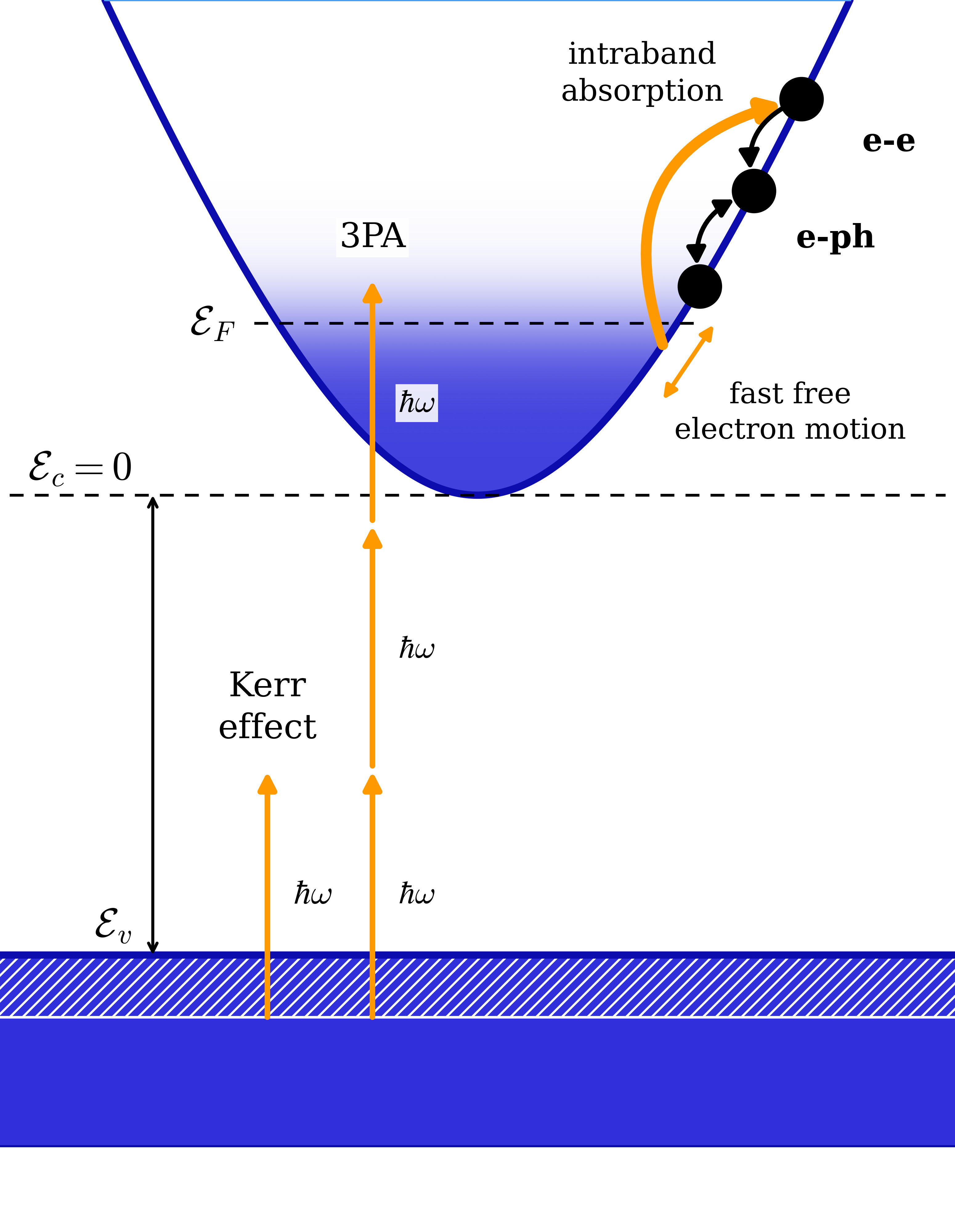}
\caption{Schematic of the electronic states and the optical processes considered in this work. The conduction band follows the Kane two-band dispersion, and the valence band is taken to be flat (dispersionless). Conduction electrons form a degenerate gas that fills the band up to the Fermi-level $\mathcal{E}_F$ (blue shading, with a thermally broadened Fermi edge). All orange arrows correspond to the photon energy $\hbar\omega$. The (thick) curved arrow marks (indirect) intraband absorption of a photon by an electron; the resulting non-thermal electron subsequently relaxes back towards the Fermi-level by (net) phonon emission (\mbox{e-ph}) and by carrier--carrier scattering (\mbox{e-e}), shown as black arrows connecting states on the dispersion curve. Filled circles denote electrons. The vertical (thin) arrows starting at the valence band represent the non-resonant single-photon virtual transition responsible for the Kerr effect, and three-photon absorption (3PA), in which three successive $\hbar\omega$ steps promote an electron into empty conduction-band states. Both processes start below the hatched dipole-forbidden region at the top of the valence band. } \label{fig:band_schematic}
\end{figure}

In this paper, based on the success of the thermal model to reproduce correctly the results of the more general non-thermal non-adiabatic coherent model~\cite{Un-Sarkar-Sivan-LEDD-II,single_cycle_nlty_Article,single_cycle_nlty_Letter}, we couple the thermal model for the electron dynamics to a rigorous FDTD (Finite Difference Time Domain) model for the Maxwell equations in a thick TCO layer. Unlike the approach in~\cite{Un-Sarkar-Sivan-LEDD-II}, the current approach addresses both field and temperature non-uniformities. We also provide a simple way to account for the temperature dependence of the permittivity within this model. To this model we add the two leading interband nonlinearities - the Kerr effect and 3 photon absorption (since two-photon absorption is well below the gap), see Fig.~\ref{fig:band_schematic} for a schematic illustration of these various processes. We then show that these two effects are far weaker than the absorptive/thermo-optic nonlinearity, thus, implying on the unlikelihood of the processes suggested so far to explain the observed behaviour for short pulses.

% Commenting out for cleanup
%{\bf this allows treating other geometries + goes beyond the (spatially-uniform) dispersive time varying media models~\cite{Paloma, Vlad, ...}}. 

%{\bf add schematic plot - inter and intra band transitions? + HHG and John's Auger?}

%\bigskip

%{\bf go into more detail?}
%\begin{itemize}
%          
%    \item thermo-optic nlty dominant. 3PA is weak and cumulative (hence, slow) + it reduces the nlty, not giving a fast feature (as implied by crude model...).
    
%    \item criticize competing mechanisms (Narimanov's Kerr; Scalora's linear absorption of HHG; Pendry's Auger; Shalaev's 3 level system with 2PA and 3PA, ...); none of them can explain so the Segev experiment; 
    
%\end{itemize}

%{\bf text on laptop elaborates the introduction...}

% For a two-column layout use \columnwidth; for one column use \linewidth.
% If the figure should span both columns, replace figure -> figure* and
% \columnwidth -> \textwidth.
% ---------------------------------------------------------------------------

\section{Model}\label{sec:model}

% {\bf need cartoon of a normally incident plane wave pump and probe on a layer, non-uniform temperature, R and T; NK - do yo have some old figure that you can edit slightly and would serve for this? \color{red} NK says - see example figure above.}
\begin{figure}
    \centering
    \includegraphics[width=0.5\linewidth]{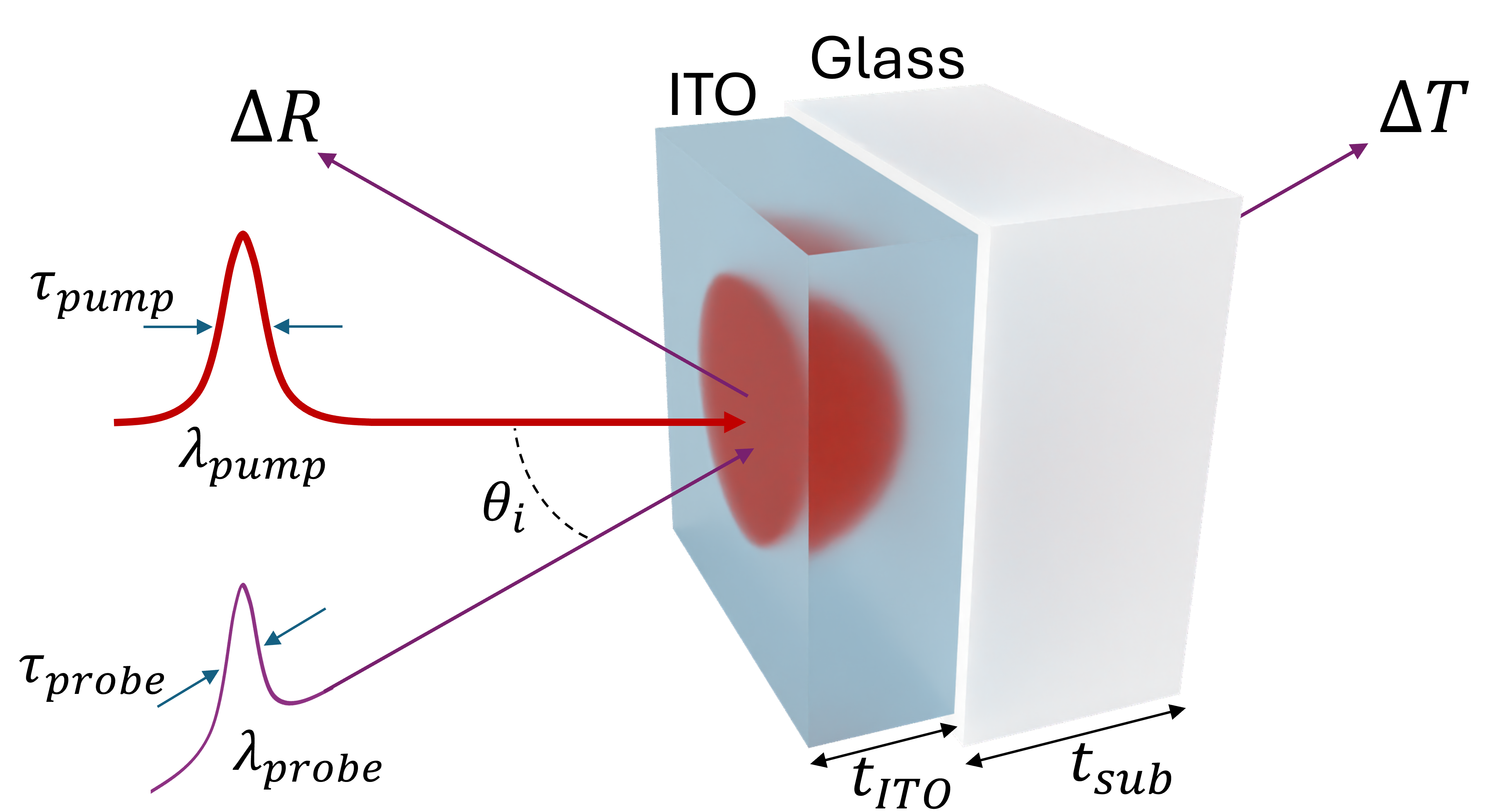}
    \caption{Schematic illustration of the Sn:In$_2$O$_3$ thin film on glass where the change in reflection $\Delta R$ and change in transmission $\Delta T$ due to the excitation of the ITO is determined. We consider a case of near normal incidence $\theta_i =0^o$ for both the pump and probe beams.}
    \label{fig:pump-probe}
\end{figure}

{\bf Configuration. } 
Given the recent interest in exploring fast features in reflection and transmission near the temporal overlap of the pump and probe pulses, as suggested by~\cite{Shalaev_Segev_nanophotonics_2023}, we focus our analysis on an Sn:In$_2$O$_3$ thin film under similar conditions as a case study. Within this context, we seek to test the hypotheses that interband interactions can give rise to these fast features by expanding the existing free-electron thermal model of the intraband nonlinearity. Specifically, we consider a $310$ nm thick ITO layer on a thick layer of glass, see Fig.~\ref{fig:pump-probe}. The (effective) plasma frequency and collision rate for the sample are set to reproduce the reported values in~\cite[Fig.~1D]{Shalaev_Segev_nanophotonics_2023} by extracting the real and imaginary permittivity values from the figure and fitting this to a Drude dispersion model seeded with properties of TCOs with a comparable crossover permittivity, see Appendix~\ref{app:linear_response}. %$\epsilon_\infty = 4$ in the ITO layer with $\omega_p = 3.07\times10^{15} rad/s$ and $\gamma = 1.29 \times 10^{14} s^{-1}$, while $\epsilon_\infty$ takes the appropriate values for air/glass in the substrate and superstrate, . 
We set a spatio-temporal Fourier-limited Gaussian pump (probe) pulse with central wavelength $\lambda_{pump} = 800$nm ($\lambda_{probe} = 1200$nm, i.e., close to the ENZ point in~\cite{Shalaev_Segev_nanophotonics_2023}). Their respective durations are set to a Full-Width-Half-Maximum of $6$ and $44$ fs, respectively. %{\bf say we limit the intensity in order not to exceed the range of temperatures...} %Pump and probe fields and currents are coupled through the common (temperature-dependent) material parameters and via the cubic (Kerr) Cross-phase modulation and the MPA.
The incident pump intensity is set to $1$ TW/cm$^2$, and the probe intensity is sufficiently low so that it does not affect the dynamics. 

% The model seeks to describe the modification of the sample's permittivity due to various nonlinear interactions, see Fig. \ref{fig:band_schematic}. Phenomenologically, we can summarize them as follows:
% \begin{equation}
% \chi^{NL} \equiv \left(\chi^{(3)}_{fc,slow} + \sum_{j=3,5,7...}\chi^{(j)}_{fc,fast} + Im\{\chi^{(5)}_{b}\} \right)_{r} + \left(\chi^{(3)}_{b}\right)_{nr}
% \end{equation}
% where the terms are grouped to illustrate resonant (r) - real state - and non-resonant (nr) - virtual state - transitions for our conditions. In the following we will outline the details of each term and its contribution to the nonlinear response of the ITO sample. Finally,}

Optical measurements of the interband absorption threshold~\cite{Hamberg1986,Kim1999,Ali2017,Klein2020} as well as recent Density Functional theory (DFT) calculations of the band structure of ITO ~\cite{DFT_ITO_DD} reveal the threshold for interband transitions is $\hbar \omega_{inter} \sim 4.5$eV; it consists of the actual $2.6$ eV gap, the Fermi energy (estimated to be $\sim 1.34$eV for the sample of~\cite{Shalaev_Segev_nanophotonics_2023}, determined through the conduction band profile of Sn:In$_2$O$_3$ see Appendix~\ref{app:linear_response}), and a range of $0.5-0.7$ eV of valence bands which are effectively dipole forbidden (so effectively inaccessible to light); the latter are not taken into account in most recent works. Thus, at least 3 photons are required to exceed the interband transition threshold of the heavily doped ITO sample; this process is depicted in Fig.~\ref{fig:band_schematic}. Higher-order multi (i.e., 4,5,...)-photon absorption events are neglected as scaling of N-photon absorption are generally progressively weaker~\cite{Brandi1983UniversalCurve}; this choice is justified a-posteriori. Note that despite the specific case taken here, the model proposed matches multiple other recent experiments~\cite{Ferrera-Scalora-ITO-2026,Sapienza_Pendry_Horsely_Tisch_2026_2}, is general and can be applied to other materials with appropriate adjustments for material and excitation parameters. Specifically, for smaller bandgap materials or higher pump energies, two-photon absorption may become relevant. % - which we neglect here since for our case study since $\hbar \omega_{pump}\approx 1.5 eV$ such that $2\hbar \omega_{pump} < E_{abs} \approx 4.5 eV$.

%\textbf{alternative we can forgo the short summary below and just dive into the details.}
%{\color{blue}The first nonlinear term is the effective $\chi^{(3)}$ arising from \textit{photo-thermal} heating of the electron temperature and subsequent \textit{thermo-optic} modulation of the permittivity through the nonparabolic conduction band~\cite{Kinsey_ENZ_OptMatExp,Khurgin_Kinsey_LPR}. The second term represents a collection of fast nonlinearities arising from free-carrier current densities within the nonparabolic conduction band~\cite{Shen2002,Khurgin_Kinsey_LPR}. They represent the free-carrier equivalent of the bound electron Kerr process. Multiple orders are intrinsically considered through the dispersion of the effective mass within the determination of the induced current density. The third-term captures the slow process of three-photon absorption (3PA), which both increases the conduction band electron density and contributes to photo-thermal heating for energy in excess of the absorption edge. The final term accounts for the conventional fast bound electron Kerr process.}

{\bf Model Description. } 
We solve Maxwell's equations in one spatial dimension and in time for a pump pulse and a delayed probe pulse normally incident on a TCO layer on a glass substrate, see Fig.~\ref{fig:pump-probe}. Specifically, we solve
\begin{eqnarray}
\frac{d}{dt}E_{pump}(z,t)  &=& \frac{1}{\epsilon_0 \epsilon^{eff}} \left(\frac{d}{dz} H_{pump}  - J_{pump}(z,t) - J_{3PA}(z,t)\right), \label{eq:E_pump} \\
\frac{d}{dt}E_{probe}(z,t) &=& \frac{1}{\epsilon_0 \epsilon^{eff}_{probe}} \left(\frac{d}{dz} H_{probe} - J_{probe}(z,t)              \right),
\end{eqnarray}
and
\begin{eqnarray}
\frac{d}{dt}H(z,t) &=& \nabla \times E(z,t) / \mu_0, \nonumber
\end{eqnarray}
where the overall electromagnetic fields are given by $E = E_{pump} + E_{probe}$ and $H = H_{pump} + H_{probe}$; $\epsilon_0$ and $\mu_0$ are the permittivity and permeability of the vacuum. An auxiliary differential equation is used to describe the electric current originating from the light-driven free (i.e., conduction band) electrons~\cite{Taflove-book}; thus, the associated (linear) Drude (i.e., intraband) current is given by
\begin{equation}\label{eq:J}
\frac{d}{dt}J(z,t) = - \gamma_D(T_e(z,t)) J(z,t) + \epsilon_0 \omega_p^2(T_e(z,t)) E(z,t),
\end{equation}
where $J = J_{pump} + J_{probe}$, $\gamma_D$ is the Drude damping rate and $\omega_p$ is the plasma frequency, both effectively averaged over electron energies (hence also $T_e$-dependent), see Appendix~\ref{app:Te_dependence}. This dependence constitutes (within the assumption of thermal equilibrium) the (slow) thermo-optic nonlinearity~\cite{Khurgin_ENZs_nlty} (to be shown to be the dominant one). By Poynting's Theorem, the absorbed electromagnetic field power density is given by
\begin{equation}\label{eq:Pabs}
P_{abs}(z,t) = E(z,t) \left(J_{pump}(z,t) + \frac{3 \hbar \omega_{pump} - \hbar \omega_{inter}}{3 \hbar \omega_{pump}} J_{3PA}(z,t)\right).
\end{equation}
where $J_{3PA}$ is the nonlinear current induced by interband 3PA transitions, see also below.

Notably, the absorption~(\ref{eq:Pabs}) is a strongly oscillatory quantity (see~\cite{single_cycle_nlty_Article}), which is nevertheless traditionally averaged out~\cite{Jackson-book}. The (positive) average gives rise to the traditional heat source. Yet, the rapid (doubled carrier frequency) oscillations represent an observable feature; specifically, they were observed experimentally in glass~\cite{Schultze_Kerr_SiO2}. They represent the fast free-carrier induced current
%(coherent) intraband nonlinear response due to conduction band (i.e., free) electrons 
discussed in~\cite{Khurgin_ENZs_nlty}. This classical feature is nothing but the (nonlinear extension) of the quiver motion / Bloch oscillations~\cite{Ashcroft-Mermin} (see Fig.~\ref{fig:band_schematic}), and was shown to resemble quantum coherent oscillation arising from quantum master equations~\cite{single_cycle_nlty_Letter,single_cycle_nlty_Article} or TD-DFT equations for electron dynamics~\cite{Schultze_Kerr_SiO2}. The  permittivity modulation associated with this oscillation cen be well approximated by a cubic (coherent) nonlinearity (and thus, incorrectly lumped onto the (by definition, incoherent) Kerr coefficient, see Appendix~\ref{app:Fast-electron}).

The absorbed power serves as a heat source that gives rise to the {\em photo-thermal} effect - it causes the electron temperature to increase, and consequently, also the temperature of the phonons to increase due to $e-ph$ collisions. The details of these processes are traditionally described by the Two Temperature Model (TTM)~\cite{Two_temp_model,delFatti_nonequilib_2000,Un-Sarkar-Sivan-LEDD-I}, i.e.,
\begin{eqnarray}
C_e(z,t) \frac{d}{dt} T_e(z,t) &=& - G_{e-ph}(z,t) (T_e - T_{ph}) + P_{abs}(z,t), \label{eq:Te} \\
C_{ph} \frac{d}{dt} T_{ph}(z,t) &=& G_{e-ph}(z,t) (T_e - T_{ph}). \label{eq:Tph}
\end{eqnarray}
Here, the electron heat capacity $C_e$ takes the standard free electron form~\cite{Ashcroft-Mermin}, with high-order corrections necessary at high temperatures (see, e.g.,~\cite[Eq.~(25)]{Un-Sivan-sensitivity})\footnote{Note that the second and fifth terms in the expansion~\cite[Eq.~(25)]{Un-Sivan-sensitivity} cancel identically. }; the electron temperature dependence of the $e-ph$ coupling relies on the analytic expression derived in~\cite{Un-Sarkar-Sivan-LEDD-I}. $C_{ph}$ is set to an empiric value of $2.4 \cdot 10^6 J / m^3 / K$; its value is secondary, as it is associated with nanosecond dynamics. 

Note that for simplicity, we do not include diffusive heat transport~\cite{ICFO_Sivan_metal_diffusion,Sivan_Spector_metal_diffusion}, as it is expected to be relatively slow due to the very short $e-e$ collision times~\cite{Un-Sarkar-Sivan-LEDD-I}\footnote{Indeed, within the Drude model, the diffusivity is linearly proportional to the $e-e$ collision time~\cite{Ashcroft-Mermin}.}. For simplicity, we also neglect ballistic transport, which could be more significant, yet, not likely to affect the dynamics in more than a mild quantitative manner, see~\cite{optic_excitation}. % }}{*not really needed and we are not sure its true.}. {\bf I would argue we kinda are; see~\cite{optic_excitation,Beraun_2006} and some unpublished calculations I am doing} \nk{}{what i'm saying this last phrase is not needed. unless we know that ballistic effects only give a mild quantitative effect, then we don't need to say it. it is not important to our result here to say this and we can just leave it at 'we don't include ballistic here'. but if you feel strongly, proceed.}

Integral expressions are given in Appendix~\ref{app:Te_dependence} also for the effective damping rate and the plasma frequency appearing in Eq.~(\ref{eq:J}); as said, these manifest the {\em thermo-optic effect} - the reverse coupling of the heat equations to the equation for the electric current, and from it to the electromagnetic fields In earlier work~\cite{Un-Sarkar-Sivan-LEDD-II}, $\omega_p$ and $\gamma_D$ (and their temperature dependence) were given implicitly within a Lindhard type integral over the (non-thermal) electron distribution. In Appendix~\ref{app:Te_dependence} we extract them into a form which requires only knowing the non-parabolicity and Fermi-level. Notably, the dependence suggested in Appendix~\ref{app:Te_dependence} generalizes the dependence of the effective mass on the non-parabolicity of the conduction band~\cite{Kinsey_ENZ_OptMatExp}. % through the dependence of the former on the electron temperature.
It also allows us to model the temperature-dependence of all these parameters up to several thousand degrees where $\gamma_D \ll \omega(_p)$, and thus, to circumvent inaccuracies occurring at such temperatures when simpler fits are used~\cite{Xian_group_ITO_2019,Boltasseva_ArXiv_2026,Sapienza_Pendry_Horsely_Tisch_2026}. 
%{\bf still i assume there ...  we limit ourselves to 5000K... otherwise, use a model with $D$ rather than a Drude model... lower than in~\cite{Un-Sarkar-Sivan-LEDD-II}}  
For the pump-probe delays studied in the current work, the dependence of the collision rate on the phonon temperature can be neglected. % the result is a change of TRA by 5e-4. {\bf within the delays studied, Tph rises by 14K, which is slightly less than the value at a few 100's of fs...} % no need to say something about the chemical potential, unless we care about the dynamics of the populations

The thermo-optic nonlinearity described so far merely implement existing knowledge in a manner that is spatially distributed (thus, improving the calculation of the spectroscopic properties) and is simpler to implement compared to early versions involving energy integrals over the electron population. We now move on beyond it, by including optical nonlinearities associated with {\em interband} transitions. %{\bf the following I would remove (as we elaborate below; either way - we do not consider linear interband absorption...) - While our focus remains on the optical response near the ENZ condition energy, interband processes can still lead to non-negligible contributions to below-gap (probe) energies, the most straightforward of which is linear interband absorption.}
%{\bf relevant to the well below gap (probe) frequencies; but could be that changes at pump/gap threshold frequencies affect other frequencies...}.}

The first additional mechanism included in our model is the instantaneous Kerr effect (i.e., the effect of {\em non}-resonant excitation of valence band electrons, see Fig.~\ref{fig:band_schematic}), which appears in the expressions for the effective permittivities: %which are given by
\begin{eqnarray}
\epsilon^{eff}_{inter,pump} = \epsilon_\infty + 3 \chi^{(3)}_b E^2_{pump}, \qquad \epsilon^{eff}_{inter,probe} = \epsilon_\infty + 3 \chi^{(3)}_b E^2_{pump}. %  \nonumber
\end{eqnarray}
Here, $\epsilon_\infty$ is the linear non-dispersive contribution to the permittivity due to interband transitions for frequencies far from the interband threshold and $\chi^{(3)}_b$ describes the instantaneous effects of self-phase modulation of the pump and cross-phase modulation of the probe\footnote{Its frequency dependence is suppressed, for simplicity.}. In Appendix~\ref{app:Kerr}, we provide a detailed estimate of the value of this coefficient, taking into account well established values known for similar doped semiconductors, and making sure not to confuse it with the intraband fast nonlinearity. We then exaggerate the value we estimated 10-fold, in order to emphasize our emerging conclusion, that the Kerr effect is much smaller than the thermo-optic nonlinearity.

A second additional nonlinear mechanism we include is (simultaneous) 3-photon absorption events which promote valence electrons to the conduction band (see Fig.~\ref{fig:band_schematic}). %\nk{Optical measurements of the interband absorption threshold~\cite{Hamberg1986,Kim1999,Ali2017,Klein2020} as well as recent Density Functional theory (DFT) calculations of the band structure of ITO ~\cite{DFT_ITO_DD} reveal that for typical pump frequencies, at least 3 photons are required to exceed the interband transition threshold of the heavily doped ITO sample, $\hbar \omega_{inter} \sim 4.5$eV; This threshold consists of the actual $2.6$ eV gap, the Fermi energy (estimated to be $\sim 1.34$eV for the sample of~\cite{Shalaev_Segev_nanophotonics_2023}), and a range of $0.5-0.7$ eV of valence bands which are effectively dipole forbidden (so effectively inaccessible to light);}{} % ...Here we focus on three-photon 3PA and neglect higher-order multi (i.e., 4,5,...)-photon absorption events as scaling of N-photon absorption are progressively weaker~\cite{Brandi1983UniversalCurve}. 
The process of 3PA results in {\em multiple} nonlinear effects, each having a distinct contribution to the dynamics; these were not previously discussed in detail before, to our knowledge. First, it gives rise to a nonlinear (bound/valence electron) current, $J_{3PA} = \beta_3 E^5 / (2 Z_0)^3$ (appearing already in Eq.~(\ref{eq:E_pump})) where $Z_0$ is the vacuum impedance and $\beta_3$ is the coefficient of three photon absorption events. This form is phenomenologic, based on the Poynting theorem; it shortcuts the more elaborate inclusion of a proper Lorentz-like equation for the current associated with the valence electrons (e.g.,~\cite{scalora_model_SHG_THG,Scalora-ITO-2020,Scalora-ITO-2025}).

% The parameter $\beta_3$ is set to $5 \cdot 10^{-27} m^3 / W^2$ {\bf Nate initially argued anything between $\beta_{3,ITO} \sim 2-5.4 \cdot 10^{-27} [m^3/W^2]$ with an elaborate reasoning why the value for ITO might be lower; and then on May 28th, argued the value should be WAY lower}, see~\cite[Table 2B]{chinese_3PA} or more systematically,~\cite{nlty_review}. 

The 3PA effect further contributes to increasing the conduction electron density (and hence, the effective plasma frequency in Eq.~(\ref{eq:J}) which is linearly proportional to it); this effect is manifested via an electric field -dependent contribution to the effective plasma frequency~\cite{Agrawal_Painter_review_OE_2007,Sivan-COPS-switching-TBG} via the electron density. Specifically,
\begin{equation}\label{eq:dN}
\frac{d}{dt} N(z,t) = \beta_3 \left(E_{pump}^2 / 2 Z_0 \right)^3 / (3 \hbar \omega_{pump}), % 4/Z0^2 converts from E to I
\end{equation}
where $\omega_{pump}$ is the pump central frequency. The decay of excess charge due to recombination typically occurs on the scale longer than the time scale of interest here ($\sim 200-300$ fs for Shockley-Reed-Hall recombination~\cite{Kinsey-Optica-2015,Clerici_Nat_Comm_2017}); for that reason, the relaxation of the electron density is neglected. % {\color{red} NK - this generally is not true for the TCOs we work with. Due to the trap density, Shockly Reed Hall recombination dominates and the relaxation time constant is typically 200-300 fs. We see this back in the early work in 2015 and again in the paper in 2017 Nat Comm where the interband excitation still has a recovery within about 1-2 ps. Yes its longer than the hot electrons but not by too much, factor of 2-3. So, this may not be the best assumption.}

Lastly, as already shown by the last term on the RHS of Eq.~(\ref{eq:Pabs}), 3PA contributes to the heating source an amount which is proportional to the excess energy deposited by the three absorbed photons above the interband transition threshold $\hbar \omega_{inter}$.

Note that as for the intraband absorption, we do not remove the fast oscillation arising from the powers of the electric field; this is yet another manifestation of the fast nonlinearity arising from the free electron response. %{\color{blue} NK-this sentence seems important given our recent discussions on the fast effect. I gather what it says but i don't get the overall point. Yes they are included but why does it matter? Can we clarify this a bit more?}
As for the Kerr effect, in Appendix~\ref{app:3PA}, we estimate the value of the 3PA coefficient based on measured values for similar doped semiconductors. The value chosen, $\beta_3 \approx 4 \times 10^{-27} m^3/W^2$, is again, most likely an exaggerated value. Thus, overall, all physical parameters associated with the inraband response are computed rigorously from microscopic models~\cite{Un-Sarkar-Sivan-LEDD-I} (and are known within reasonable sample-to-sample variation) whereas in the absence of verified DFT calculations for the nonlinear optical coefficients of ITO, the missing values are estimates based on the measured response of comparable wide bandgap semiconductors.

All quantities are computed on the same time grid and advanced using a 4th-order Runge-Kutta method, whereas the electric and magnetic field are staggered in space (as for the Yee grid); the currents are spatially-collocated with the electric field. We have verified that in the linear regime, the probe dynamics is in agreement with the reflection and transmission Fresnel theory (with the relative error being no more than a fraction of a percent).

The model described above exceeds the state-of-the-art for TCOs (\cite{Scalora-ITO-2025,Ferrera-Scalora-ITO-2026}) by relying on rigorous temperature dynamics (extracted from microscopic electronic modelling~\cite{Un-Sarkar-Sivan-LEDD-I,Un-Sarkar-Sivan-LEDD-II}) and corresponding temperature-dependence of optical and thermal parameters as well as on their unique non-parabolicity and low electron density; this contrasts the phenomenological fit of the nonlinearity via a series of high-order powers of the electric field~\cite{Scalora-ITO-2020,Scalora-ITO-2025,Ferrera-Scalora-ITO-2026,Purdue-oscilatory_nlty_ITO-2026}). Our model also adds interband MPA that to our understanding, was never accounted for before in a complete manner in TCO studies. Our model nevertheless does not include coherent nonlinearities giving rise to nonlinear wave mixing, and in particular, not the more sophisticated nonlinearities usually included in hydrodynamic models~\cite{scalora_model_SHG_THG,Ciraci:2012_PRBR,Maxim_hydrodynamic,Scalora-ITO-2020,Scalora-ITO-2025}, as those too involve mostly coherent second-order nonlinearities, which are out of the scope of the current study. %{\bf + weak and/or limited to the surface; + do not enjoy the ENZ/vg enhancement; in contrast, Scalora/Marcello point out they are relatively strong (and even base their interband effect on the absorption of HHGs), but NK agreed that they do not affect the main dynamics + the experiment shows a delayed response that is not captured by scalora's HHG.} {\color{red} NK - I think the point here is that we focus on the index change so we dont' care about fast wave mixing terms. We only need the Kerr process. nothing in chi2 is useful to us. The key point is that we therefore do not consider absorption of generated harmonics and its potential effect upon carrier density, as Scalora and Marcello point out. I don't believe this story, but cannot rule it out entirely at this point. This would be the main thing worth noting.} 
Similarly, as for~\cite{single_cycle_nlty_Letter,single_cycle_nlty_Article}, our model does not account for the strong nonlocality in TCOs~\cite{Scalora-ITO-2020,Thouin2025}, although it may become relevant as the film becomes thinner. % \sout{because only for very thin {\bf replace by thinner? Mike claims non-locality is particularly strong for ITO; cannot say anything intelligent about this claim, so I am ok, with giving him credit here.} films.} % {\color{red} NK says - since we do not have a threshold for when this matters or not, perhaps we just avoid saying a particular value altogether. we just say we don't include it here but it may be relevant for some cases [reference to mike]}

Furthermore, our model does not account for stimulated emission (which reduces the total absorption) or excited state absorption (i.e., non-simultaneous multi photon absorption events), because those effects seem to mostly cancel each other~\cite{single_cycle_nlty_Letter}.

Finally, we do not include the Kerr nonlinearity of the substrate and focus our analysis on the thin ITO film. Isolation of the thin film response in experiment can be achieved under cross-polarized excitation between pump and probe, where the effect of the substrate is sufficiently small compared to the TCO that it can be neglected in most scenarios. However, under parallel polarization this may not always be true~\cite{Benis2022ENZ,Sapienza_Pendry_Horsely_Tisch_2026} and care must be taken in analysis. Yet in this case, proper normalization can remove this effect for well-known substrates such as fused silica to isolate the thin film response. % {\color{red} NK - yes this is usually the case. It can only be relevant under || polarization of pump-probe. If you use cross-pol, then it is more or less negligible. We measure it just to make sure but its usually so small we don't care. But for || pol, this isn't the case, see S. Benis et al "Extremely large nondegenerate nonlinear index and phase shift in epsilon-near-zero materials [Invited]", Fig. 2. Substrate matters. I've edited the text here to add this point.}% not simulated (because GVD in substrate not accounted for). % our simulations do not capture this

% for sufficiently long wavelength, $\epsilon_\infty \sim \omega_{p,inter}^2(N(t))/\omega_{inter}^2(T_e(t))$

% 18/12/25 - removed ADE for Kerr (PJ_inter_pump) and associated predictor-corrector iterations (which anyhow did not work well). replaced by an effective (dispersion-free) permittivity. very stable.

\section{Results}
\begin{figure}[h]
\centering
\includegraphics[width=1\columnwidth]{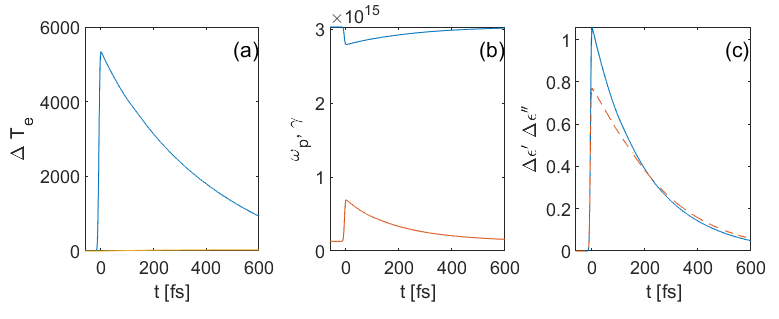}
\caption{Dynamics of spatially-averaged pump-induced change of (a) electron temperature $T_e$, (b) effective plasma frequency $\omega_p$ (blue) and collision rate $\gamma_D$ (red), and (c) real (blue) and imaginary (dashed red) part of the intraband part of the ITO layer permittivity $\epsilon'$, at the central probe frequency $\omega_{probe}$ for incident intensity of $1$ TW/cm$^2$; the interband effects of Kerr and 3PA are not included in this simulation. % (d) The {\bf interband part of the} ITO layer permittivity $\epsilon'_{Kerr}$~(\ref{eq:Delta_R} -- edit) {\bf add imaginary part? --- I am showing WAY too many orders of magnitude} 
} \label{fig:param_dynamics}
\end{figure}

% {\bf General nonlinear dynamics.} 
Overall, the pulses traverse the layer, with the pump being partially absorbed, so that heat, nonlinear currents and excess free carriers are being generated in the ITO layer, along with cross-phase modulation being induced upon the probe by the Kerr effect. The dynamic permittivity changes are primarily generated by the thermo-optic nonlinear response which themselves arise from the increase of the electron and phonon temperatures. Fig.~\ref{fig:param_dynamics} shows the resulting {\em spatially-averaged} dynamics of the electron temperature and various material properties (over the thickness)%should we plot the spatio-temporal maps? the point is that they ain't accurate because I do not account for dynamics due to diffusion (nor ballistic transport), but this seems not to matter so much, I carefully say...}
. One can see that the change builds up on the pump-duration time scale, but relaxes on a much longer time scale, dictated primarily by $e-ph$ coupling; notably, there is no fast oscillatory feature. For our choice of parameters, the electron temperature rises by several thousand degrees, the effective plasma frequency drops by $\sim 20 \%$, and the the electron collision rate grows drastically. 

As observed for longer pulses before~\cite{Un-Sarkar-Sivan-LEDD-II}%{\color{red} NK - We never measure permittivity directly, always R,T or similar and infer permittivity change. If you want a direct measurement of permittivity the earliest would be \cite{Shalaev_Faccio_NLO_ENZ}, we also give permittivity values here \cite{Kinsey-Optica-2020} although not the focus of the paper}
, the real part of the permittivity undergoes order unity changes, and the increase of the imaginary part\footnote{Note the difference with respect to the theory in~\cite{Lee-Marini} which predicted that the collision rate should decrease upon heating. That prediction originated from the unjustified neglect of the role of non-thermal holes (i.e., classical collision dynamics).} are only somewhat smaller % THIS AWAITS IW's input - {\bf compare to longer pulses and lower intensity in Viewpoint + IW's response!;} 
% one can see a fast feature (discussed with Alu at some point); it is more prominent for higher intensities, but it does not seem to affect the spectroscopic quantities -- potentially because it is not very strong + too brief}.{\color{blue} NK says - I'm not sure what this is about so can remove from my view.} % {\bf so simulate higher intensities? for a longer pulse at 1TWcm2, this feature is more pronounced. but Te goes to 25K, and I am not sure that stimulated emission would not modify the behaviour significantly... }

In Fig.~\ref{fig:Kerr_3PA_effect}(a), we show the dependence of the spectroscopic quantities ($T$, $R$, $A$) on the pump-probe delay time. % The linear (i.e., pre-pump) transmissivity $T$, reflectivity $R$ and absorptivity $A$ are $\sim 45\%$, $\sim 15\%$, and $\sim 40\%$, respectively. % describe resulting temperature distribution - higher near interfaces..}
Upon illumination, these quantities change by $\sim 10-30\%$. % $\sim -17\%$, $\sim -10\%$, and $\sim +25\%$, respectively; these are $\sim -37\%$, $\sim -66\%$, and $\sim +62\%$ {\em relative} changes. 
% {\bf NK - what do you think about the next sentence? - 
Note that since $A$ depends on the delay between the pump and probe, one can expect the peaks of  and reflectivity not to be at the same position (as pointed out already in~\cite{Scalora-ITO-2025}). %-- but these are not clearly seen in the latest round of results...} {\color{red} NK - well, i'd probably just show the R,T,A curves rather than list the values. Then this sentence would make more sense. }

We now turn to consider the secondary MPA and Kerr interband effects. We will see that they are negligible in comparison to the thermo-optic nonlinearity.

% \begin{figure}[h]
% \centering
% \includegraphics[width=1\columnwidth]{param_dynamics.png}
% \caption{Dynamics of spatially-averaged pump-induced change of (a) electron temperature $T_e$, (b) plasma frequency $\omega_p$ (blue) and collision rate $\gamma$ (red), and (c) real (blue) and imaginary (red) part of the ITO layer permittivity $\epsilon'$, at the central probe frequency $\omega_{probe}$ for a $1$ TW/cm$^2$; the interband effects of Kerr and 3PA are not included in this simulation. {\bf add intensities here? } } \label{fig:param_dynamics}
% \end{figure}

\subsection{Kerr nonlinearity}
When we include the effect of the Kerr nonlinearity in the ITO layer, a clear instantaneous change is observed on the background of the thermo-optic nonlinearity (Fig.~\ref{fig:Kerr_3PA_effect}(b)). Following Appendix~\ref{app:Kerr}, we set $\chi^{(3)}_b = 10^{-19} m^2/V^2$ so that the corresponding maximal index change ($\Delta \epsilon_{Kerr} \equiv 6 \chi_b^{(3)} E^2_{pump} \sim 3 \cdot 10^{-19} (1.37 \cdot 10^9)^2 \sim 0.5$) would be comparable to the thermo-optic nonlinearity (see Fig.~\ref{fig:param_dynamics}(c)). Despite that, we observe transmissivity (reflectivity) changes of only $\sim 1.65\%$ ($\sim 0.21\%$) with respect to the case with only a thermo-optic nonlinearity, i.e., $\sim 5$-fold ($\sim 50$-fold) smaller changes. % {\color{red} NK - I am not following what is stated here. If the change in permittivity was set to be about the same as the thermo-optic portion, e.g. about 1, how is the percentage value so much lower? Can we show this graphically and then refer to a plot to illustrate the point? For example, breakdown all of the components and their contributions to delta epsilon to see relative magnitudes?}
Thus, our rigorous treatment shows that $\Delta R \ll \Delta \epsilon_{Kerr}$. This is a result of the transient nature of the permittivity change, which limits the interaction time that leads to reflection~\cite{Khurgin-review-2023}; indeed, this result contrasts the CW illumination case for which $\Delta R \sim \Delta \epsilon_{Kerr}${\color{blue}~\cite{Narimanov_2025}}. 
%This was overlooked in the analysis of (who also considered a comparable permittivity modulation strength) {\bf NK - maybe we should rephrase this...}{\color{red} NK - I wouldn't point it out directly. You can just leave the first sentence.} 
Since the scaling of the Kerr-induced changes are linear with the incident intensity (not shown), the emerging conclusion is that in order to explain the experimental results of~\cite{Shalaev_Segev_nanophotonics_2023}, the ITO Kerr coefficient should be \emph{more than} an order of magnitude higher than the already 10-fold exaggerated value set in our simulations to play a meaningful role in $\Delta R, \Delta T, \Delta A$ and explain the rapid transient reported in~\cite{Shalaev_Segev_nanophotonics_2023}.

% {\bf note that the Kerr effect has also a non-instantaneous indirect impact - it affects the absorption level during the pulse presence, and hence, the differential transmissivity/reflectivity have a small non-zero value at long times.} {\color{blue} Is this relevant? If not lets leave it out. Story is already complex}

\subsection{Multi-photon absorption}

\begin{figure}[h]
\centering
\includegraphics[width=1\columnwidth]{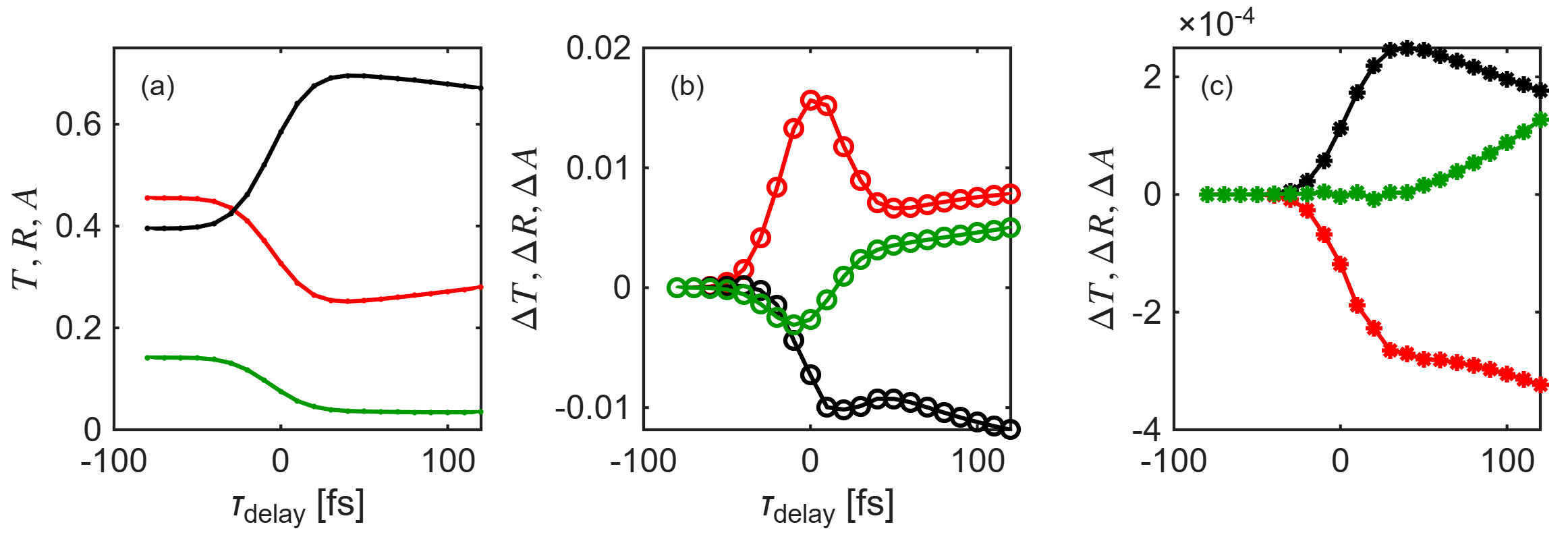}
\caption{(a) Transmissivity (red), Reflectivity (green) and Absorptivity (black) through an ITO layer with a pure thermo-optic nonlinearity (dots) illuminated by a pump pulse of maximal intensity of $1$ TW/cm$^2$ as a function of the pump-probe delay. (b) and (c) show the change in these quantities due to a(n exaggerated) (b) Kerr (circles) and (c) 3PA (stars) nonlinearities, respectively. } \label{fig:Kerr_3PA_effect}
\end{figure}

Fig.~\ref{fig:Kerr_3PA_effect}(c) shows the effect of 3PA on the transmissivity, reflectivity and absorptivity. Overall, for our generic choice of parameters, 3PA hardly affects these quantities - its effect is $O(10^{-4})$, i.e., even smaller than the Kerr effect. The main reason for the smallness of 3PA is the fact that for the experimental configuration we are reproducing~\cite{Shalaev_Segev_nanophotonics_2023} (which is representative of many other experimental systems~\cite{Ferrera-Scalora-ITO-2026,Sapienza_Pendry_Horsely_Tisch_2026_2}), 3PA is at the spectral threshold for interband transitions. 

\begin{figure}[h]
\centering
\includegraphics[width=1\columnwidth]{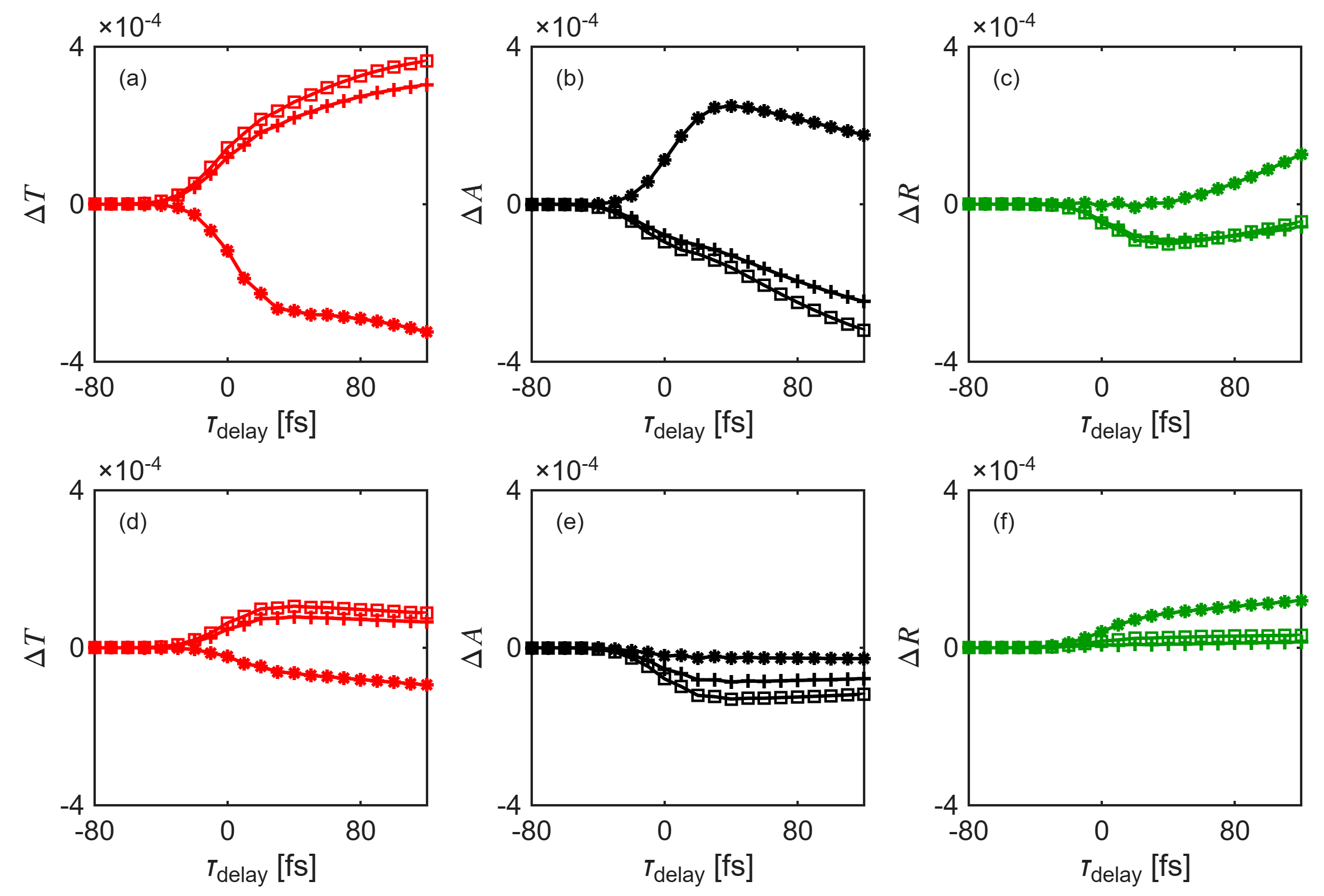}
\caption{Breakdown of the effects caused by 3PA for incident intensity of $1$ TW/cm$^2$. The change of transmissivity due to the 3PA induced current (last term on the RHS of Eq.~(\ref{eq:E_pump})) is shown by squares. The addition of the 3PA-induced heat generation (last term on the RHS of Eq.~(\ref{eq:Pabs}); $+$ symbols) causes a slight reduction of the transmissivity change, and the addition of the excess charge carriers to the conduction band (see Eq.~(\ref{eq:dN}); stars, as in Fig.~\ref{fig:Kerr_3PA_effect}(c)) inverts the sign of the change. Similar behaviour is seen for the absorptivity (except for opposite signs); the reflectivity is affected mildly. (d)-(f) shows the same data for incident intensity of $0.5$ TW/cm$^2$. } \label{fig:3PA_breakdown}
\end{figure}

An additional reason is more subtle and somewhat unexpected. We observe that the different effects caused by 3PA have opposing contributions to the permittivity, thus, partially canceling each other. In particular, Fig.~\ref{fig:3PA_breakdown}(a) shows that for $1$ TW/cm$^2$ incident intensity, while the nonlinear current $J_{3PA}$ leads to an increase of the transmittance, the generation of excess free carriers in the conduction band (see Eq.~(\ref{eq:dN})) has a slightly stronger effect of opposite sign (heat generation also has a negative effect, but is much weaker); the effects on the absorptivity have opposite signs (Fig.~\ref{fig:3PA_breakdown}(b)), and the reflectivity is less sensitive to either (Fig.~\ref{fig:3PA_breakdown}(c)). This cancellation reduces the overall effect of 3PA by another order of magnitude. To add to the complexity, similar results for half the incident intensity reveal that the reflectivity and absorptivity show different trends (e.g., the differential absorption actually changes sign) and while the transmissivity seems to be qualitatively similar and quantitatively lower, it is only 4-fold and not 8-fold lower, as naively expected. This tells us that one cannot predict reliably what will be the magnitude of 3PA for much higher incident intensities. This aspect is therefore missed in studies that do not consider all three effects arising from the multi-photon absorption events~\cite{Purdue-oscilatory_nlty_ITO-2026}.

% \sout{One may claim that a stronger 3PA effect would be achievable with a higher pump frequency, but this was not attempted experimentally so far, to our knowledge} {\bf Marcello's paper reaches 10 TW/cm2, no?}{\color{blue} several works now do. I'd just leave this out becuase we can't simualte in the current case anyway since it would get the CB electrons too hot. If i recall correctly, most of the recent works start to see the effect around 1 TW/cm2, and since we don't see it i think the argument still holds. However if pushed we can note that it may be an open question.}. {\color{blue}
While 3PA may be stronger for higher intensities and/or other TCOs which have a somewhat narrower gap, it is to be emphasized that, like the thermo-optic nonlinearity, all three effects arising from 3PA are cumulative over time (i.e., non-instantaneous), and hence, they are unlikely to yield a fast feature as seen in~\cite{Shalaev_Segev_nanophotonics_2023}. %This is rooted in the increased free carrier generation and heat associated with 3PA, which dominate the (seemingly instantanous) nonlinear current. Furthermore, it weakens the nonlinearity - it makes the transmissivity and absorptivity smaller compared to the pure thermo-optic nonlinearity case. In that sense, in contrast to the desire to achieve a strong and fast nonlinearity, 3PA is slow causes the nonlinearity to be weaker, like stimulated emission~\cite{single_cycle_nlty_Letter}; yet, unlike the latter effect, 
Furthermore, even if a fast feature does arise, 3PA actually pushes the material system further away from thermal equilibrium, thus making the realization of temporal photonic crystals even less likely.

Our results shed light on recent studies that associated fast features to MPA. Specifically,~\cite{Purdue-oscilatory_nlty_ITO-2026} claimed that 2PA and 3PA may explain the rapid (and oscillatory) behaviour of the transmissivity and reflectivity changes induced by a few femtosecond pump pulse absorbed in a thin ITO layer. Notably, however, the model in that work did not consider the dominant thermo-optic nonlinearity, and instead, relied on the phenomenologic MPA effects within a 3 level system model (rather than within the continuous band characteristic of solids). %\sout{such a model is overly simplistic, and the nonlinear mechanism cannot provide the instantaneous behaviour needed to observe the fast feature}. {\color{red} NK says - this was stated earlier in this section paragraph "Further, our work demonstrates..."}. 
Furthermore, the interband threshold was likely underestimated, leading to the unlikelihood that 2PA processes contribute meaningfully to the response even in the presence of an empty conduction band. % \sout{but also the fact that 2PA processes are impossible for the spectral configuration in the experiment even in the presence of an emptied conduction band was overlooked.} 
% \nk{parameters with incorrect units - their 2PA coeff is $10^{-12} cm^4 sec$ - units must be wrong...?? ... + they tracked the population dynamics partially and crudely; furthermore, it is to be noted that the probe duration in that work was relatively long (100 fs), so that it is difficult to associate with the measurements faster dynamic changes.}{remove if you're okay YS}

A related paper~\cite{Ferrera-Scalora-ITO-2026} used the model described in~\cite{Scalora-ITO-2025} to claim that the intense illumination of light with the ITO layer causes the generation of high (5th, ...) harmonics, which are then linearly absorbed due to single photon interband transitions. Here too, the thermo-optic nonlinearity is replaced by a somewhat phenomenological polynomial expansion of instantaneous nonlinearities. Our results on the smallness of the Kerr and MPA effect and the ratio between them challenges the likelihood of high-order harmonic process becoming significant as they rely on similar nonlinear susceptibilities which should not vary by orders magnitude, and raises the need for an explicit quantification of the HHG efficiency~\cite{Yang-HHG-CdO-natphys-2019}. More generally, the temporally delayed response of the effect is key to the correct interpretation of the reported results. Specifically, it is observed experimentally that the contribution ascribed to the harmonic absorption occurs at times beyond the pulse width of the pump, yet the modulation of the permittivity due to harmonic generation and its absorption is expected to reach its maximum near the peak of the pump pulse. The delay in the observed feature suggests an effect with a longer rise time / cumulative effect, rather than harmonic absorption.

% {\color{red} NK says - I propose to remove. if you would like to keep this i would rather argue on the grounds of Miller's rule. We can estimate chi3 and chi5 and it should be much weaker than they require.} However, as far as we understand, the authors do not quantify what is the intensity of the generated harmonics, and what portion of it gets absorbed and re-emitted % the following is my misunderstanding --- {\bf NK - you wanna justify this, bc Scalora states differently - 5PA is important for him??} {\bf I don't understand how they argue that 5PA is so important. it is against most convention so i can't really say much other than this is not usually the case. It comes from their modeling. As you know they do not predict anything, but simply fit coefficients to measured data so it does not necessarily have to be true.}

Finally, both these studies did not account for the nonlinear current $J_{3PA}$ nor to the heating associated with it, and hence, not to the mutual cancellations.

% \begin{figure}[h]
% \includegraphics[width=1\columnwidth]{3PA_effect_breakdown.png}
% \caption{Same as in Fig.~\ref{fig:3PA_effect}, showing the change in the spectroscopic quantities due to each individual contribution to 3PA - the nonlinear current ($>$), the excess heat generation (o) and the increase of the conduction electron density for each quantity, respectively. } \label{fig:3PA_effect_breakdown}
% \end{figure}

% This shows that models based on MPA alone can be considered as faithful representation the underlying physics unless they consider all three opposing effects. {\color{red} NK - i wouldn't directly say this - its too combative. I would note it earlier on as an advancement of the technique.}

\section{Conclusions}

% {\color{blue} NK - Does the planned journal have a conclusion section or just a final discussion section? If so i might start the discussion here and leave the two paragraphs above for the conclusion.}

In this work, we advanced our understanding of the nonlinear optical response of TCOs to short and intense pulse illumination. Our work relies on the success of thermal models to reproduce the dynamics observed from the more complex non-thermal models. First, we
% provides an extension of the dispersive time modulated models used in~\cite{paloma,Purdue, ...} to a system with spatial dependence, as well as coupling to temperature and permittivity dynamics. It also 
provided a simplification of the electronic models for the dependence of the TCO permittivity on the electron and phonon temperatures developed in~\cite{Un-Sarkar-Sivan-LEDD-I,Un-Sarkar-Sivan-LEDD-II,single_cycle_nlty_Article,single_cycle_nlty_Letter}. This simplification enables easy implementation in future studies. Second, unlike these previous works, we implement the model for a layer geometry which supports nonuniform fields and temperatures. 

% First, as shown for ITO particles in~\cite{single_cycle_nlty_Letter,single_cycle_nlty_Article}, we show that the behaviour identified for pulses of a few 10s of femtosecond or longer persists also for single cycle pulses

% not sure this is needed, nor corrct (maybe this is the reason for the comparable changes of the real and imaginary part?) {\bf ??} Overall, the nonlinearity (in terms of relative permittivity changes or the relative change in the spectroscopic quantities) is qualitatively and quantitatively similar to that observed for on-resonance excitation by longer pulses in previous work on ITO layers and particles~\cite{Un-Sarkar-Sivan-LEDD-II}. % at $1 TW/cm^2$ and $6$ fs pump (so incident fluence of $60 J / m^2$) compared to, say, $75 GW/cm^2$ and $220$ fs AT RESONANCE which gave a fluence of $165 J / m^2$ and permittivity changes of 1.5 / 15 fold; so comparable...}

Our work demonstrated the dominance of the intraband (thermal) nonlinearity of TCOs over the contribution of valence to conduction band {\em virtual} transitions (i.e., Kerr nonlinearity) and {\em real} transitions (due to multi photon absorption); the former give rise to a small (3rd order) instantaneous permittivity change, and the latter give rise to an even smaller (5th order) accumulation of a permittivity change that rises as rapidly as the pulse duration (hence, the intraband nonlinearity), but decays on a near-picosecond time scale. These findings are robust to variations of the optical and thermo-physical (heat capacities, $e-ph$ coupling etc.) parameters, see discussion in Appendix~\ref{app:Te_dependence}. Thus, it seems that these effects cannot explain the rapid decrease of transmissivity observed in~\cite{Shalaev_Segev_nanophotonics_2023}.
% with lower $\e_F$ - linear response changes, but non-linear trends remain similar...

This finding has two main consequences. First, it shows that realization of temporal photonic crystals is unlikely in ITO (and TCOs in general) because of the dominant (and cumulative nature of the) thermo-optic nonlinearity - surpassing the damage threshold is avoidable only for very low repetition rates.

Second, this demonstrates the quantitative unlikelihood of several mechanistic explanations proposed for the fast spectroscopic features observed in~\cite{Shalaev_Segev_nanophotonics_2023} and recent works that associated dynamic changes measured from TCO layers to the interband effects.

Additional nonlinear effects associated with interband transitions (such as the AC Stark effect~\cite{Autler_Townes}, depletion of the valence band population, Fermi smearing and band shifting~\cite{Rosei_Au_diel_function_temp}, a change of the interband collision rate etc.) affect primarily the UV spectral range, i.e., close to the interband absorption edge. While the current work does not rule out the importance of these effects, they are not expected to be significant at the near IR frequencies associated with the ENZ point and their inclusion may be at the focus of future efforts. % \sout{A rigorous study of such effect should be performed once reliable DFT data would become available~\cite{DD_interband...}.}}
In this vein, one should emphasize that our approach for the interband nonlinearities should be made more accurate for higher frequencies, for the high harmonics generated in the system, and especially for pump frequencies reaching the interband absorption threshold~\cite{Clerici_Nat_Comm_2017}. For example, one should then adopt a full Lorentzian approach that is more suitable to treat short pulses, chirp effects, and resonant enhancement of the nonlinear response; this was doe in~\cite{Scalora-ITO-2020,Scalora-ITO-2025,Ferrera-Scalora-ITO-2026}.

% {\color{blue} NK - is this worth mentioning? i would imaging this would be an even weaker additional effect} We also do not include thermal expansion, stress and strain since those occur on the slow time scale related with the (much smaller) rise of the phonon temperature). % can be unjustifiably incorporated into the Kerr coefficient when interpreting experimental measurements.

Our model opens the way to the consideration of other explanations of the yet unexplained fast features observed in~\cite{Shalaev_Segev_nanophotonics_2023}, in particular, in more extreme conditions of pulse energies, as those studied in~\cite{Sapienza_Pendry_Horsely_Tisch_2026}, both of which require a more advanced model that captures the very high temperatures developing in these systems. This advance will be needed in order to test the claim 
% 3PA quenches Kerr; i prefer to ignore this...
% \begin{itemize}
%  {\bf RE-THINK THIS --- tripled linear loss affected the dynamics only in a modest quantitative manner, as did a change of the non-parabolicity. }
in~\cite{Sapienza_Pendry_Horsely_Tisch_2026}, the fast feature seen in experiments was ascribed to impact ionization (aka carrier multiplication); a simple numerical implementation showed a good match to the experimental data. However, the inverse effect which was not accounted for, sometimes called Auger recombination can be easily shown to be of identical probability, if the electron system is in thermal equilibrium. While earlier work relied upon in the current study showed that this is a good approximation, those studies did not include the dynamics of the valence band population. Thus, the question of which of the two effects (sometimes referred to jointly as Auger scattering) dominate would require adding the valence band electrons to the non-thermal model~\cite{single_cycle_nlty_Article}. While the current work provided a step forward in that direction, this task still lies beyond the current scope .

% \item {\bf remaining explanations of fast features - Gunn (intra-conduction band), ballistic transport. 

% \item refer to Tisch's claim on interband gamma increase; explain why N3PA cannot be too strong (Pendry)}.
% \end{itemize}

% {\bf NK - what about Marcello's argument about circular polarization? ---> i suppose the email correspondence from aug. 29-30 implies it is not such a clear cut argument?

% should we not say something about the claims in  M. Clerici, N. Kinsey, C. DeVault, J. Kim, E. G. Carnemolla, L. Caspani, A. Shaltout, D. Faccio, V. Shalaev, A. Boltasseva, M. Ferrera, Nat. Commun. 2017, 8 1–7. -- it involves interband nlties which clearly were stronger. or the effect there was different? the cancelling seems not to happen there}

% Isolating fast intraband nlty from fast interband nlty and slow intraband nlty in experiments measuring nonlinear index change is challenging as the fast {\bf / instantaneous} response is generally expected to be weaker than the slow (thermal) nonlinearity by 10-100$\times$ in TCOs~\cite{Khurgin_Kinsey_LPR} {\bf refer to discussion in appendix. or in the discussion section?} % {\color{red} NK - lets not include this yet. this would be good content for the opinion or similar piece. Or perhaps you can cite your most recent paper on Nanophotonics?}

% say here that the dependence on the intensity is sublinear; as for longer pulses~\cite{Un-Sarkar-Sivan-LEDD-II}, hence, not shown + that this MAY seem like saturation, but it ain't

% \section{Conclusions}

\bigskip

% {\bf Acknowledgments.} The authors wish to thank M. Scalora and M. Ferrera for multiple in depth discussions on the various underlying effects and ways to model them.

\appendix

%\section{TBC gain:}
%$$
%\int_{-\infty}^t e(-(t'/\tau_p)^2) e^{-(t - t')/\tau} e^{- i \delta t'} dt' =
%\frac{\sqrt{\pi}\,\tau_p}{2}
%e^{- \frac{t}{\tau} + \frac{\tau_p^2}{4}\left(\frac{1}{\tau}-i\delta\right)^2}
%\left[1 + \operatorname{erf}\left(
%\frac{t - \frac{\tau_p^2}{2}\left(\frac{1}{\tau}-i\delta\right)}{\tau_p}
%\right)
%\right].
%$$
%... the field is not a true Gaussian, due to interaction with the layer, but never mind...

\section{Determination of material parameters}
\subsection{Linear room temperature response}\label{app:linear_response}
The linear properties of the ITO film were determined through fitting of the linear permittivity provided by~\cite{Shalaev_Segev_nanophotonics_2023}. Due to the narrow spectral range provided, data for ITO films available to the authors with a comparable permittivity crossover and loss were used to solidify parameters. Since we only require the long-wavelength response of the ITO film (beyond 1100nm), the interband transitions can safely be ignored and lumped into a single $\epsilon_\infty$ within the Drude model
\begin{equation}
\epsilon(\omega;T_e,T_{ph}) = \epsilon_\infty - \frac{\omega_p^2(T_e)}{\omega^2 + i \omega \gamma_D(T_e,T_{ph})}. \label{eq:Drude}
\end{equation}
Here, $q_e$ is the electron charge, $\epsilon_0$ is the vacuum permittivity, $m_e^{Kane}$ is the electron mass appearing in the Kane formula for the non-parabolic band structure.

Within this limit, we fit the standard Drude model to the ITO permittivity of~\cite{Shalaev_Segev_nanophotonics_2023} and find $\epsilon_\infty = 3.99, \omega_p = 3.07 \times 10^{15} rad/s, \gamma_D = 1.29 \times 10^{14} s^{-1}$. These values agree with ellipsometry results for similar ITO films. With the Drude properties known, we can then leverage the Kane-band model of the conduction band energy profile to uniquely determine the Fermi-level of the material for a given scattering rate. Taking:
\begin{equation}
\e = \sqrt{\left(\frac{1}{2C}\right)^2 + \frac{\hbar^2}{2C m^{Kane}_e}k^2} -\frac{1}{2C}
\end{equation}
with $m^{Kane}_e = 0.2 m_e$ and $C = 0.3497 eV^{-1}$, we gradually increase the Fermi energy and evaluate the carrier density and average effective mass using the definition from~\cite{Kinsey_OME_2020}, comparing the resistivity and plasma frequency to the value extracted from the linear permittivity fit. The result identifies $\e_F = 1.34$ eV, $n_e = 1.13 \times 10^{21} cm^{-3}$ and $m^{Drude}_e = 0.388 m_e$. The final value of resistivity matches that expected from the Drude-model fit of the permittivity, and the generated permittivity subsequently provides a good fit to the linear permittivity dispersion of the sample provided by~\cite{Shalaev_Segev_nanophotonics_2023}.

% @Yonatan, I can add figures of the fit here if you would like but i don't think its necessary. -- agreed...

\subsection{Nonlinear high temperature response}\label{app:Te_dependence}
In order to feed the simulations, we would now like to derive an analytic form for the thermo-optic response of the ITO as a function of the extracted values for the non-parabolicity $C$ and the Fermi energy $\e_F$, in a manner that reconciles the intraband Lindhard formula~\cite[Eq.~(9)]{Un-Sarkar-Sivan-LEDD-II}\footnote{Note that that early expression (i.e., in~\cite{Un-Sarkar-Sivan-LEDD-II}) failed to distinguish between the Kane mass $m_e^{Kane}$ and the temperature dependent effective mass entering the Drude formula, $m_e^*$. }
\begin{equation}
\varepsilon(\omega;T_e,T_{ph}) = \varepsilon_\infty - \frac{q_e^2}{\varepsilon_0 m^{Kane}_e} \int_0^\infty \frac{\rho_e(\E;C) \kappa(\E;C) f(\E)}{(\omega + i \eta(\E;T_e,T_{ph}))^2} d\E, \label{eq:Drude_Lindhard}
\end{equation}
% {\bf Claude wrote this as $v^2$?} 
with the Drude formula~(\ref{eq:Drude}). Here,
\begin{equation}
\rho_e(\E;C) = \frac{1}{2\pi^2} \left(\frac{2 m^{Kane}_e}{\hbar^2}\right)^{3/2} \sqrt{\E(1 + C \E)} (1 + 2 C \E), %\qquad \rho_0 = \frac{1}{2\pi^2} \left(\frac{2 m^{Kane}_e}{\hbar^2}\right)^{3/2}, \label{eq:rho0}
\end{equation}
is the density of electron states,
\begin{equation}
\kappa(\E;C) = \frac{1 + \frac{8}{3}C \E (1 + C \E)}{(1 + 2 C \E)^3}, \nonumber
\end{equation}
is a correction factor associated with the non-parabolicity, and
\begin{equation}
\eta(\E;T_e,T_{ph}) = A T_{ph} + U K_e (\pi k_B T_e)^2 + U K_e (\E - \e_F)^2, \label{eq:eta}
\end{equation}
is the electron collision rate contribution to the permittivity. Here, we use the (standard) energy-dependent (Fermi liquid based) expression for $e-e$ collision rate, with $K_e$ given by~\cite[Eq.~(19)]{Un-Sarkar-Sivan-LEDD-I} and $U$ being the Umklapp branching ratio; below, it is set to unity since its value is anyhow embedded in $K_e$. For simplicity, we neglect the seemingly opposing negative correction $\sim (\e - \mu(T_e) T_e$) obtained in~\cite{single_cycle_nlty_Letter}, and the positive correction obtained in~\cite[Eq.~(18)]{Un-Sarkar-Sivan-LEDD-I} due to the non-parabolicity. As for noble metals, the dependence on the phonon temperature is assumed linear due to its relatively modest and slow increase, but this dependence can be generalized if necessary. % $A$ and $G_{e-ph}$ can be connected (see discussion with Claude):  $G \approx ​\pi^2 k_B^2 ​n_e​​ \omega_D​ A / 2 \e_F $.

Using the expression for $\mu(T_e,C,\e_F)$ in~\cite[Eq.~(24)]{Un-Sarkar-Sivan-LEDD-I}) and neglecting the deviation from thermal equilibrium, we insert the Fermi function
$$
f(\E) \to f^T(\E;T_e,C,\e_F) = \frac{1}{1 + \exp\left(\frac{\E - \mu(T_e,C)}{k_B T_e} \right)} \nonumber
$$
into the Lindhard integral~(\ref{eq:Drude_Lindhard}). In the small-damping regime, $\gamma_D \ll \omega$, the integral term in this equation, responsible for the free electron response, is approximated by
%\begin{equation}
%\frac{1}{(\omega + i \gamma)^2} = \frac{1}{\omega^2} \left[1 - 2 i \frac{\gamma}{\omega} - 3 \left( \frac{\gamma}{\omega} \right)^2 + \mathcal{O}\left( \frac{\gamma^3}{\omega^3} \right)\right]. \nonumber % \label{eq:den-expansion}
%\end{equation}
%Retaining the leading nontrivial orders and substituting into Eq.~\eqref{eq:Drude_Lindhard} yields
\begin{equation}
I(T_e,T_{ph},C) \equiv \int_0^\infty \frac{\rho_e(\E,C)\kappa(\E,C) f^T(\E,T_e)} {(\omega + i \eta(\E,T_e,T_{ph}))^2} d\E \approx \frac{1}{\omega^2} I_0(T_e,C) - i \frac{2}{\omega^3} I_1(T_e,T_{ph},C). \nonumber
\end{equation}
Here, we defined the frequency-{\em in}dependent quantities
\begin{eqnarray}\label{eq:I01-def}
I_0(T_e,C,\e_F) &\equiv& \int_0^\infty \rho_e(\E;C) \kappa(\E;C) f^T(\E;T_e,C,\e_F) d\E, \\
I_1(T_e,T_{ph},C) &\equiv& \int_0^\infty \rho_e(\E;C) \kappa(\E;C) f^T(\E,T_e,C,\e_F) \eta(\E;T_e,T_{ph}) d\E.
\end{eqnarray}
%Thus,
%\begin{equation}
%\Re[I] \cong \frac{I_0}{\omega^2}, \qquad \Im[I] \cong -\,\frac{2 I_1}{\omega^3}.
%\label{eq:ReImI}
%\end{equation}

\begin{figure}[h]
\centering
\includegraphics[width=1\columnwidth]{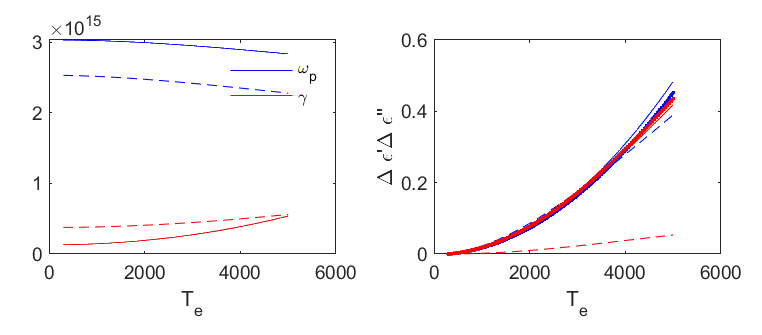}
\caption{(a) Electron temperature dependence of the effective plasma frequency (blue) and collision rate (red) computed via Eqs.~(\ref{eq:eff-params}) (solid lines) or as %via the formulae in Naama Harcavi's Thesis (equivalent to
in~\cite{Un-Harcavi-Sivan-LEDD-Viewpoint} %)
(dashed lines). (b) Electron temperature dependence of the real (blue) and imaginary (red) parts of the Drude permittivity computed via the exact Lindhard integral (Eq.~(\ref{eq:Drude_Lindhard})) (dots) or via the approximations based on Eqs.~(\ref{eq:eff-params}) (solid lines) and based on the parameters in%Naama Harcavi's Thesis
~\cite{Un-Harcavi-Sivan-LEDD-Viewpoint} (dashed lines).} \label{fig:expansion}
\end{figure}

Matching to the small damping limit of the Drude formula~(\ref{eq:Drude}), $\omega_p^2 / [\omega (\omega + i \gamma_D)] = \omega_p^2\left(\omega^{-2} - i \gamma_D\omega^{-3} + \cdots\right)$ allows identifying the effective plasma frequency and collision rate as
\begin{equation}
\omega_p^2(T_e;C,\e_F) = \frac{q_e^2}{\varepsilon_0 m_e^{Kane}} I_0(T_e,C,\e_F), \qquad \gamma_D(T_e,T_{ph}) = 2 \frac{I_1(T_e,T_{ph},C,\e_F)}{I_0(T_e,C,\e_F)}.
\label{eq:eff-params}
\end{equation}
These parameters are plotted in Fig.~\ref{fig:expansion}(a) as a function of the electron temperature, using their room temperature values extracted in Appendix~\ref{app:linear_response}.

Note that the temperature dependence of the (effective) plasma frequency is equivalent to a temperature-dependent effective mass used in some previous work~\cite{Kinsey_ENZ_OptMatExp,Khurgin_Kinsey_LPR}. Also note the factor 2 that originates from the approximate denominator appearing in the (energy-space) Lindhard form~(\ref{eq:Drude_Lindhard}).

Now, we use the values of $\epsilon_\infty$, $\omega_p$ and $\gamma_D$ extracted from the measurements at room temperature (see Appendix~\ref{app:linear_response}) to determine consistently the values for $C$, $A$, $K_e$ and $\e_F(n_e)$. % $A$ is not exactly Geph

In principle, one can now proceed by applying the Sommerfeld expansion to $I_0$ and $I_1$. However, this expansion is divergent, so that it is bound to fail above a threshold electron temperature (for which $k_B T_e$ is no longer smaller than $\mu(T_e)$); in particular, since $\e_F$ here is relatively low compared to a noble metal, the electron temperature rise is significant, and $\mu$ monotonically decreases with $T_e$~\cite{Un-Sarkar-Sivan-LEDD-II}, the range of validity of the Sommerfeld expansion is only $\approx 2000-3000$K. Instead, we compute $I_0(T_e,C,\e_F)$ and $I_1(T_e,T_{ph},C,\e_F)$ numerically, and then interpolate the resulting expressions and feed them to the FDTD code. This approach would also eventually fail once $\gamma_D$ becomes comparable to $\omega_p$, as expected to happen at very high temperatures~\cite{Un-Harcavi-Sivan-LEDD-Viewpoint}.

Fig.~\ref{fig:expansion}(b) demonstrates the good match between the approximated calculations (Eqs.~(\ref{eq:eff-params})) and the complete Lindhard expression~(\ref{eq:Drude_Lindhard}). However, it also shows that the resulting expression for ($\gamma_D$, hence,) $\epsilon''$ there is far greater sensitivity to the electron temperature compared to the calculation in~\cite{Un-Harcavi-Sivan-LEDD-Viewpoint}. The reason for that is the different proportions given to the $e-ph$ and $e-e$ contributions to $\eta$~(\ref{eq:eta}). This stems from the unknown Umklapp weight for TCOs, and the difficulty to separate these contributions to the permittivity. This unavoidable arbitrariness has a quantitative effect on the strength of the thermo-optic nonlinearity.

\section{Estimate of magnitude of interband nonlinear coefficients}
\subsection{Estimate of the Kerr coefficient} \label{app:Kerr}
There are several theoretical approaches to estimate the value of the Kerr coefficient. The bound electronic effect far from resonance can be derived from quantum mechanical perturbation theory~\cite{Boyd-book,Khurgin_Kinsey_LPR,Narimanov_2025} or through the well developed scaling theory for the nonlinear refractive index of semiconductors~\cite{non-deg-2PA-CREOL}. Here, we presume that the virtual transitions of valence band electrons in ITO are similar to those that occur in the undoped In$_2$O$_3$ material at the same frequency. We consider the role of doping upon the interband process through its shift of the Fermi-level, but consider the electrons as heavy so that their contribution to the linear permittivity in the near-infrared is negligible. In this approximation, we leverage experimental data of lightly doped ZnO and ZnS \cite{nlty_review,chinese_3PA} - wide bandgap semiconductors with similar optical and electrical properties. In the wavelength range of 750nm to 950nm (comparable to that of the pump wavelength in our case), the average value of the intensity dependent refractive index is $n_2 \approx 6.8\times10^{-6}$cm$^2$/GW. Assuming negligible two-photon absorption, reasonable given that $2\omega_{pump} < \omega_g$, this corresponds to value of $\chi^{(3)}_b\approx 1\times10^{-20}$ m$^2$/V$^2$~\cite{delCoso2004Relation}. We note that this value likely represents an overestimate of the $\chi^{(3)}_b$ for ITO as the bandgap of heavily doped ITO is approximately 1eV larger than undoped ZnO and ZnS, and the probe wavelength of 1225nm used in~\cite{Shalaev_Segev_nanophotonics_2023} is longer than the dataset - both of which are expected to result in a reduced value of intensity dependent index. {\em Despite this}, in subsequent calculations we exaggerate further the bound electron nonlinearity by 10-fold and take $\chi^{(3)}_b\approx 1\times10^{-19}$ m$^2$/V$^2$. This places its strength on par with the thermo-optic nonlinearity (see Appendix~\ref{app:Fast-electron}) to more readily test its role in generating a fast response.

The above discussion involved valence to conduction band virtual transitions. In~\cite{Narimanov_2025}, Narimanov considered inter-conduction band virtual transitions. While such virtual transitions between occupied and unoccupied conduction-band states may contribute to the nonlinear optical response, their impact is expected to be limited in degenerately doped TCOs such as ITO. Away from conduction-band critical points, the relevant conduction bands are highly dispersive, resulting in a relatively small joint density of states (JDOS) and a broad distribution of transition energies. Consequently, any enhancement associated with inter-conduction band transitions is expected to be spectrally diffuse rather than resonant. Although localized regions of large JDOS could produce stronger effects, such features are not expected for the nearly linear conduction-band dispersion near the Fermi level in ITO. We therefore expect inter conduction band processes to provide, at most, a secondary contribution relative to other established nonlinear mechanisms. % {\bf specifically, valence to conduction virtual transitions?  those two types can be distinguished experimentally - via polarization or different gap energies...

\subsection{Estimate of the 3PA coefficient} \label{app:3PA}
The value of the 3PA coefficient of ITO was estimated by averaging experimental measurements of from similar wide-bandgap semiconductors (ZnO, ZnS,~\cite{nlty_review,chinese_3PA}) under similar experimental parameters, as well as by fitting this data with the multi-photon absorption model by Brandi and de Arajo~\cite{Brandi1983UniversalCurve} and then adjusting for the excitation parameters and bandgap of the ITO film. Averaging the measured 3PA coefficients near 800nm, we find $\beta_3 \approx 4 \times 10^{-27} m^3/W^2$ ($Im\{\chi^{(5)}\} \approx 4.8 \times 10^{-39} m^4/V^4$~\cite{NLO_Handbook}, assuming $n_4$ is negligible) valid under an excitation regime where $\hbar \omega_{pump}/\E_g = 0.43$. However, for ITO with a bandgap of $4.5$eV excited at $800$nm, the band-edge normalized excitation energy $\hbar \omega_{pump}/\E_g$ is $0.34$, which is significantly smaller than the experimental measurements of ZnO and ZnS. Due to the strong dependence upon the pump photon energy in this energy region, we estimate the scaled value for ITO to be $\beta_3 \approx 7 \times 10^{-29} m^3/W^2$ ({$Im\{\chi^{(5)}\} \approx 8.5 \times 10^{-41} m^4/V^4$})~\cite{Brandi1983UniversalCurve}. The average of the measured ZnO and ZnS data therefore represents a high-end value of 3PA while the scaled value represents a more realistic estimate of the expected 3PA the ITO case. Despite this, we employ only the former (high end) values, to account for the broad spectral bandwidth of the pump.

\section{Connection to Fast Free Electron Nonlinearity}
\label{app:Fast-electron}
Here, we would like to provide a note on the connection between previous descriptions of the fast free electron nonlinearity~\cite{Khurgin_Kinsey_LPR} and the absorptive/thermo-optic nonlinearity included in Section~\ref{sec:model}. From the perspective of~\cite{Khurgin_Kinsey_LPR}, the absorptive/thermo-optic nonlinearity contains two simultaneous contributions - a slow nonlinearity that arises from free-carrier absorption of the pump energy, and a fast nonlinearity that arises from induced currents within the non-parabolic conduction band \cite{Shen2002,Khurgin_Kinsey_LPR}. The former has been called the intraband nonlinearity~\cite{Clerici_Nat_Comm_2017}, and is hallmarked by the fact that the resulting modulation lasts for times beyond the pump temporal width, typically with relaxation times on the scale of 100's fs. The latter is a near instantaneous nonlinearity of free-electron motion that exists only during the pump pulse. Although this distinction is useful to clarify terminology, in the approach outlined here, both contributions are captured by solely determining the induced current density through the analytical description of the momentum/energy-dependent effective mass.

A final point of clarification is that while the fast free-carrier nonlinearity is nearly instant, and also results in permittivity modulation, it is distinct from the bound electron Kerr effect and they should not be confused. These two effects can have different magnitudes and dependence upon incident intensity. As an example, the \textit{effective} third-order susceptibility of the fast free-carrier nonlinearity in ITO was predicted to be $\chi^{(3)}_{eff} \approx 2.5 \times 10^{-19}$ m$^2$/V$^2$~\cite{Khurgin_Kinsey_LPR}, while the bound electron Kerr process herein is estimated to be $\chi^{(3)}_{b} \approx 1 \times 10^{-20}$ m$^2$/V$^2$.

%We can summarize this effect from the frequency-domain perspective as outlined in~\cite{Khurgin_Kinsey_LPR} as:
%\begin{equation}
%J = - q_e \sum_k v_z(k,t) = - q \left( v_z(k_{z0}) + \frac{\partial v_z}{\partial k_z} \delta k(t) + \frac{1}{2}\frac{\partial^2 v_z}{\partial k_z^2}\delta k^2(t)+ \frac{1}{6}\frac{\partial^3 v_z}{\partial k_z^3}\delta k^3(t)+...\right),
%\end{equation}
%where $v_z$ is the velocity along the direction of the field, $k_{z0}$ is the initial momentum of the electron, and $\delta k$ is the shift in momentum due to the applied field.
%Since $\delta k \propto E$, the current is effectively a power-series expansion in the applied field, allowing the introduction of conventional susceptibilities $\chi^{(j)}$. These susceptibilities result in permittivity modulation similar to the {\em bound} electron Kerr process.
%As an example, the third-order term in ITO was predicted to be $\chi^{(3)}_{eff} \approx 2.5 \times 10^{-19}$ m$^2$/V$^2$~\cite{Khurgin_Kinsey_LPR} while even order terms cancel since the conduction band is nearly symmetric about its energetic minimum. This is crucial because the effective-Kerr arising from fast free electrons response appears qualitatively similar in experiment to a true $\chi^{(3)}_{b}$ yet arises from the motion of {\em free} electrons between real states where scattering and loss can occur - and as such, they can contribute to the absorbed pump power.

\end{document}